\documentclass[aps,preprint]{revtex4}
\usepackage{ifpdf}
\ifpdf
\usepackage[pdftex]{graphicx}
\usepackage{epstopdf}
\else
\usepackage{graphicx}
\fi
\usepackage[nolist]{acronym}
\usepackage{float}
\usepackage{calc}
\usepackage[nomessages]{fp}
\usepackage{subfig}
\restylefloat{table}
\newcommand{\pureW}{$\alpha$-W}
\newcommand{\OOW}{O$_2$-W}
\newcommand{\NNW}{N$_2$-W}
\begin{acronym}

\acro{mbe}[MBE]{molecular beam epitaxy}
\acro{afm}[AFM]{atomic force microscopy}
\acro{aes}[AES]{auger electron spectroscopy}
\acro{rbs}[RBS]{Rutherford backscattering spectrometry}
\acro{xrr}[XRR]{X-ray reflectometry}
\acro{xps}[XPS]{X-ray photoelectron spectrometry}
\acro{XRD}{X-ray diffraction}
\acro{arxps}[ARXPS]{angled-resolved X-ray photoelectron spectrometry}
\acro{rms}[RMS]{root mean square}
\acro{uhv}[UHV]{ultra high vacuum}
\acro{ald}[ALD]{atomic layer deposition}
\acro{beem}[BEEM]{ballistic electron emission microscopy}
\acro{stm}[STM]{scanning tunneling microscopy}
\acro{UHV}{ultra high vacuum}
\acro{SIMS}{secondary ion mass spectroscopy}
\acro{TEM}{transmission electron microscopy}
\acro{FCC}{face centered cubic}
\acro{BCC}{body centered cubic}
\acro{SHE}{Spin Hall effect}
\acro{GSHE}{giant spin Hall effect}
\acro{SHA}{spin Hall angle}

\end{acronym}

\begin{document}

\title{Weak dependence of phase fraction and dopant concentration in $\beta$, $\alpha$  phase mixture of W}

\author{Avyaya J. Narasimham}
\email{ajayanthinarasimham@albany.edu}
\affiliation{College of Nanoscale Science and Engineering,
University at Albany, SUNY, Albany, New York 12203, USA}

\author{Farheen N Sayed}
\affiliation{Department of Materials Science and NanoEngineering, Rice University ,Houston, Texas 77005-1827, USA}

\author{Praneet Adusumilli}
\affiliation{IBM Research , Albany, New York 12203, USA}

\author{Steve Novak}
\author{Vincent P.~LaBella}
\email{vlabella@sunypoly.edu}
\affiliation{Colleges of Nanoscale Science and Engineering, SUNY Polytechnic Institute, Albany, New York 12203, USA}

\acresetall
\date{\today}

\begin{abstract}

Dopant concentration and $\alpha$, $\beta$ phase fraction is studied in doped W thin films. Oxygen doped $\beta$ W films are found to have 9.0~$\pm$~0.9 at.\% of oxygen as measured from \ac{SIMS}. Whereas, nitrogen doped $\beta$ W films have 0.15~$\pm$~0.02 at.\% of nitrogen, much lower than the theoretically predicted value of 11 at.\%. $\beta$-W films partially phase transform to $\alpha$ by annealing at 175~$^{\circ}$C, up to maximum time of 72 hours. Weak dependence between dopant concentration and phase fraction is observed in the annealed films. The growth exponent of the phase transformation was calculated to be $<$1 in both the films, indicating a non-uniform nucleation of the $\alpha$ phase.
\end{abstract}

\pacs{}

\keywords{}

\maketitle

\acresetall

$\beta$ W films are known to exhibit \ac{GSHE} which can be utilized to manipulate the magnetic moment of an adjacent ferromagnetic layer~\cite{pai:apl:101:122404, hao:apl:106:182403, demasius:ncom:7:10644, neumann:apl:109:142405, zhang:apl:109:192405-1, he:apl:109:202404}. The ferromagnetic layer can be the free layer, in a 3-terminal magnetic tunnel junction device~\cite{liu:sc:336:555} or a charge-coupled spin logic device~\cite{datta:apl:101:252411}. \ac{SHE} generates a spin current density ($\vec{j}_s$) as a response to charge current density ($\vec{j}_c$), which are related by $\vec{j}_s \propto (\vec{S} \times \vec{j}_c$), where $\vec{S}$ is the spin polarization of electron~\cite{hankiewicz:prb:70:241301(R)}. The effect is quantified by a \ac{SHA}, defined as $\vec{J}_s/\vec{J}_c$~\cite{hirsch:prl:83:1834, hoffmann:ietm:49:10, liu:sc:336:555, dyakonov:pla:35A:459}. Films with higher \ac{SHA} require lower operating $\vec{J}_c$, thus consuming less power~\cite{kent:natn:10:187}. In a recent work by Deamsius~\textit{et al.}, highly oxygen doped $\beta$ W films have the highest \ac{SHA} of  0.5, to date~\cite{demasius:ncom:7:10644}. It has been demonstrated experimentally that the meta-stable $\beta$ phase tungsten has a higher \ac{SHA} and higher resistance than $\alpha$ phase tungsten~\cite{pai:apl:101:122404, hao:apl:106:182403}.

Earlier studies have reported a $\beta$ to $\alpha$ phase transformation ($\beta\rightarrow\alpha$) in W, induced by thickness, local heating, and diffusion of dopants~\cite{rossnagel:jvstb:20:2047, weerasekera:apl:64:3231,okeefe:jap:79:9134, karabacak:tsf:493:293}. Producing highly stable $\beta$ W films can be challenging since it transforms to $\alpha$ phase as the thickness exceeds about 5~nm~\cite{vullers:tsf:577:26,shen:jap:87:177, shen:msea:284:176, kizuka:jcg:131:439, okeefe:jap:79:9134, karabacak:tsf:493:293, rossnagel:jvstb:20:2047, aouadi:jvsta:10:273, petroff:jap:45:2545}.
It has been experimentally demonstrated that the presence of dopants during the growth stabilizes and promotes $\beta$ phase over $\alpha$, where films much thicker than 5~nm have been achieved~\cite{shen:msea:284:176, weerasekera:apl:64:3231, narasimham:aipa:5:117107, basavaiah:apl:12:259, liu:acma:104:223, karabacak:tsf:493:293}. \textit{Ab initio} calculations have shown that, the interstitial sites in tetrahedrally close-packed structures, when occupied by either nitrogen, oxygen, or fluorine stabilize the $\beta$ phase of tungsten at concentrations of 7--11~at.\%~\cite{sluiter:prb:80:220102}.  Theoretical studies of such first order structural phase transitions in materials have demonstrated that they are strongly influenced by the strength and nature of the correlation of nuclei~\cite{rickman:pre:95:022121,tong:jcp:114:915,rickman:acma:45:1153}. In some recent physical vapor deposition studies of tungsten, it is observed that doping N$_2$ or O$_2$ gas during deposition favors the growth of $\beta$ phase over $\alpha$~\cite{liu:acma:104:223, narasimham:aipa:4:117139,narasimham:aipa:5:117107}. It is also been shown experimentally that  higher concentration of these dopants, during growth, skew the phase fraction towards $\beta$ W~\cite{liu:acma:104:223,karabacak:tsf:493:293,okeefe:jap:79:9134}. Typical operating $\vec{J}_c$ and resistivity values for \ac{SHE} devices are in the order of 10$^6$~A~cm$^{-2}$ and 300~$\mu \Omega$-cm, respectively~\cite{pai:apl:101:122404, hao:apl:106:182403}. Such large currents in highly resistive metals can lead to Joule heating during multiple read and write operations~\cite{bi:apl:105:022407, conte:apl:105:122404, sethi:apl:107:192401}. Such devices are required to operate at 180~$^\circ$C or above for industrial and automotive applications~\cite{kent:natn:10:187}. These factors can induce the phase transformation, $\beta\rightarrow\alpha$ tungsten and thus degrade the device performance over time. This motivates a need to study the thermal stability of these doped $\beta$ W films.

In this letter, $\beta$ W films are grown either with a continuous flow of O$_2$ gas or a periodic pulse of N$_2$ gas~\cite{narasimham:aipa:4:117139,narasimham:aipa:5:117107}. Annealing in \ac{UHV} environment at a temperature of 175$^\circ$C, for different time lengths, up to 72 hours, was carried out on these films to study the phase transformation of $\beta\rightarrow\alpha$~W. \ac{XRD} was utilized to determine the relative phase fractions of $\beta$ and $\alpha$ and estimate their lattice parameters by using Rietvield refinement. To determine dopant concentration \ac{SIMS} depth profiling was performed on each of the annealed films. In as-deposited condition a marked reduction in dopant concentration was found in the films grown using pulsed N$_2$ gas. No strong correlation between dopant concentration and phase fraction was observed in annealed films. Avrami growth exponent of the $\beta\rightarrow\alpha$  transformation in nitrogen doped tungsten and oxygen doped tungsten films was estimated to be $<1$, suggesting a similar transformation for both the films.


Three different 20-nm-thick tungsten samples were fabricated on a 300~mm Si(001) substrate in a physical vapor deposition system (Singulus-PVD).  Different conditions were utilized during the W deposition for the three samples: (1) deposition of W (\pureW), (2) deposition of W while introducing a 2~sccm of O$_2$ gas (\OOW), and (3) deposition of a 5~nm SiN layer followed by the W deposition while introducing a 1~sccm N$_2$ gas pulse with a period of 2 seconds (\NNW). Each W film was then capped with 50-nm-thick Ni layer to avoid oxidation of the W films. The \OOW\ and \NNW\ films were carefully cleaved along $<$110$>$ direction into several individual samples. These samples were subjected to annealing in a \ac{UHV} environment at a temperature of 175$^{\circ}$C for different durations ranging from 1~hour to 72~hours. When the anneal time ended, the sample was quickly transferred to a cooler (room temperature) \ac{UHV} environment and then removed from \ac{UHV} after 60 seconds.

After annealing, grazing incidence \ac{XRD} was performed using a Bruker X-ray diffractometer in a parallel-beam geometry with a fixed 0.5$^{\circ}$ incidence angle (sealed Cu source). \ac{SIMS} depth profile was acquired on a Physical Electronics Model 6650 dynamic \ac{SIMS} instrument. \ac{SIMS} depth profiles were acquired using 3~keV Cs bombardment at 60$^{\circ}$ incidence under \ac{UHV} conditions. For oxygen measurements, negative ions were detected and for nitrogen measurements CsN$^+$ ions were detected. The depth scale was calibrated using layer thicknesses determined by \ac{TEM} (not shown). The oxygen and nitrogen concentrations were calibrated using ion implanted tungsten standards.

\begin{figure}[h!]
\includegraphics[width=7cm]{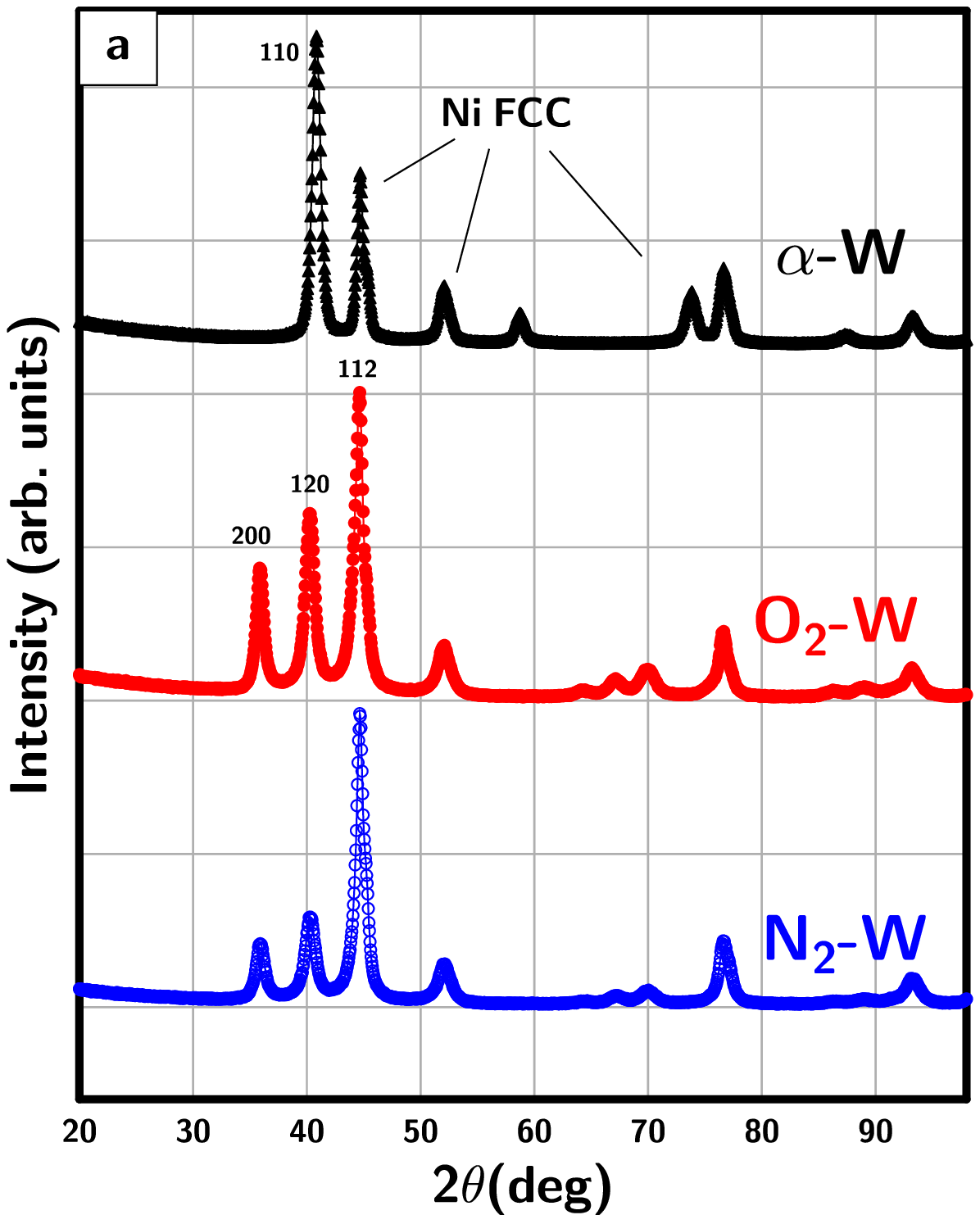}
\includegraphics[width=7cm]{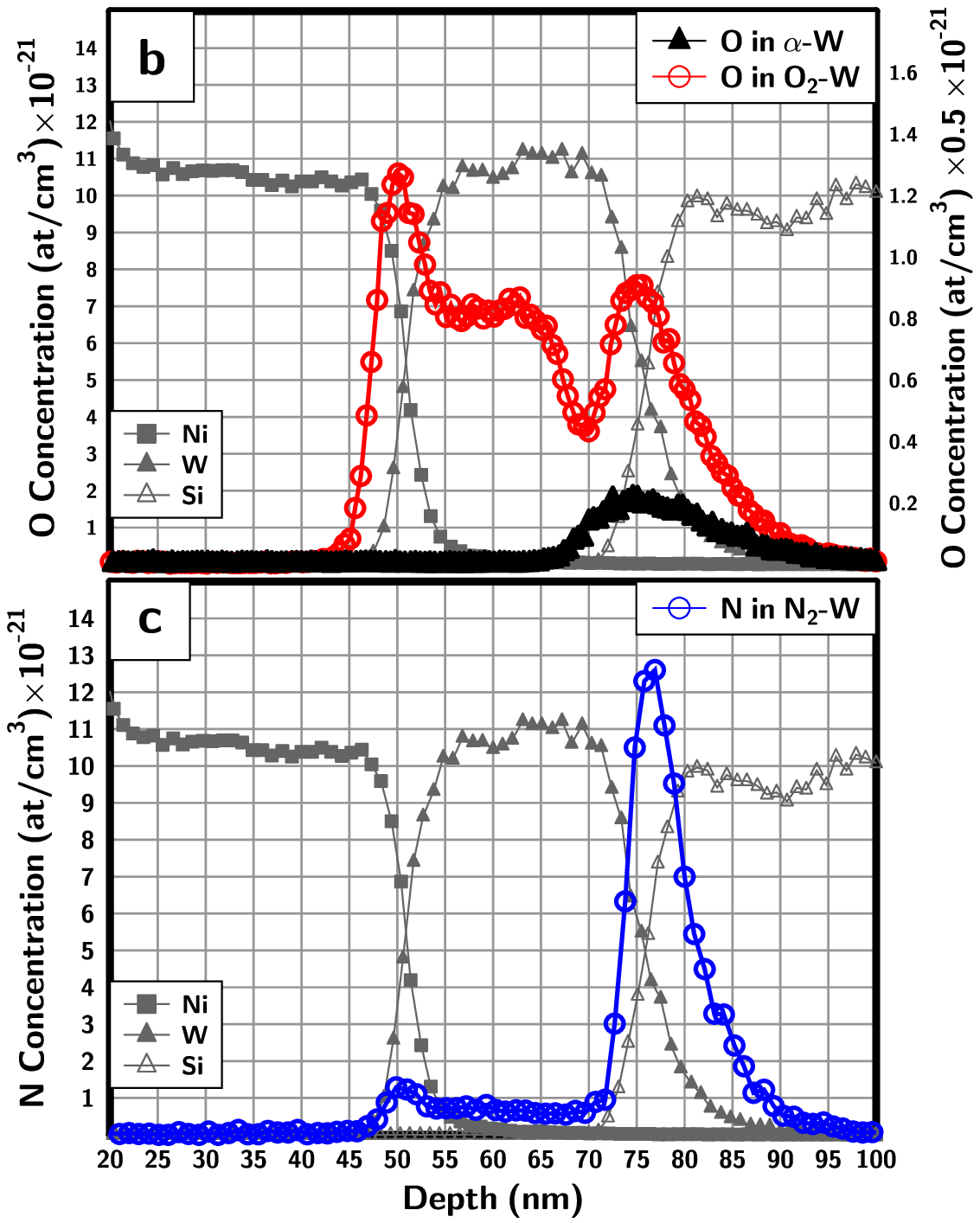}
\caption{(Color online) (a)Grazing incidence \ac{XRD} of the as deposited 20 nm thick tungsten films of \pureW\ (filled triangle), \OOW\  (filled circle) and  \NNW\ (empty circle). An offset in the patterns is added for ease of distinction (b)\ac{SIMS} depth profile of as deposited \OOW\ and \pureW\ films (on right axis) is shown. (c)\ac{SIMS} depth profile of as deposited \NNW\ film is shown. Normalized concentrations of nickel, tungsten and silicon are plotted.}
\label{f1}
\end{figure}

The grazing incidence \ac{XRD} intensity versus 2$\theta$ scans are displayed in Fig.~\ref{f1}a. The \ac{FCC} peaks of Ni overlap with $\beta$ W peaks as indicated in the figure. Using Rietveld refinement it is determined that lattice parameters are 3.17~\AA\ for the \pureW\ sample, 5.04~\AA\ for the \OOW\ sample, and 5.03~\AA\ for the \NNW\ sample. A 100~\% $\beta$ phase is estimated for \OOW\ and \NNW\ films.

The \ac{SIMS} depth profile of oxygen and nitrogen in the tungsten films are displayed in Fig.~\ref{f1}b and c. The concentration in the plots are calibrated using ion implanted standards of oxygen and nitrogen in a W film. For the \pureW\ film only 0.03~at.\% of oxygen was measured in the tungsten layer, which is close to the instrumentation background (not shown). In the \OOW\ film about $9.0\pm0.9$~at.\% of oxygen was measured in the tungsten layer. In the \NNW\ film no measurable oxygen was detected and $0.15\pm0.02$ at. \% of nitrogen was measured in the tungsten layer. Nickel, tungsten and silicon concentrations are normalized and plotted to indicate the different layers in the film.

\begin{figure}[h!]
\includegraphics[width=7.5cm]{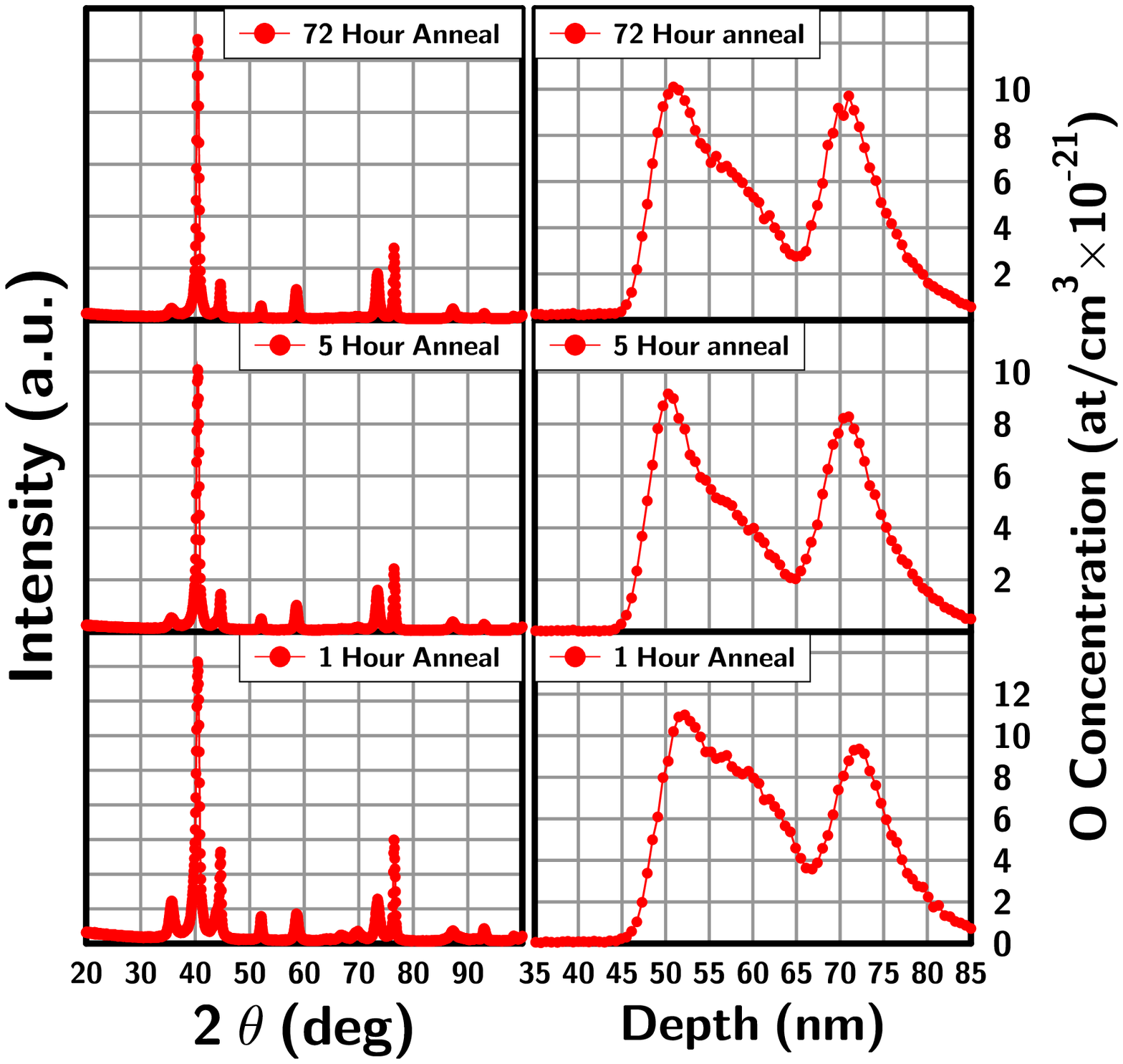}
\caption{(Color online) Grazing incidence \ac{XRD} (Left) and \ac{SIMS} (Right) depth profile of \OOW\ film annealed for 1, 5 and 72 hours is plotted.}
\label{f2}
\end{figure}

\begin{figure}[h!]
\includegraphics[width=7.5cm]{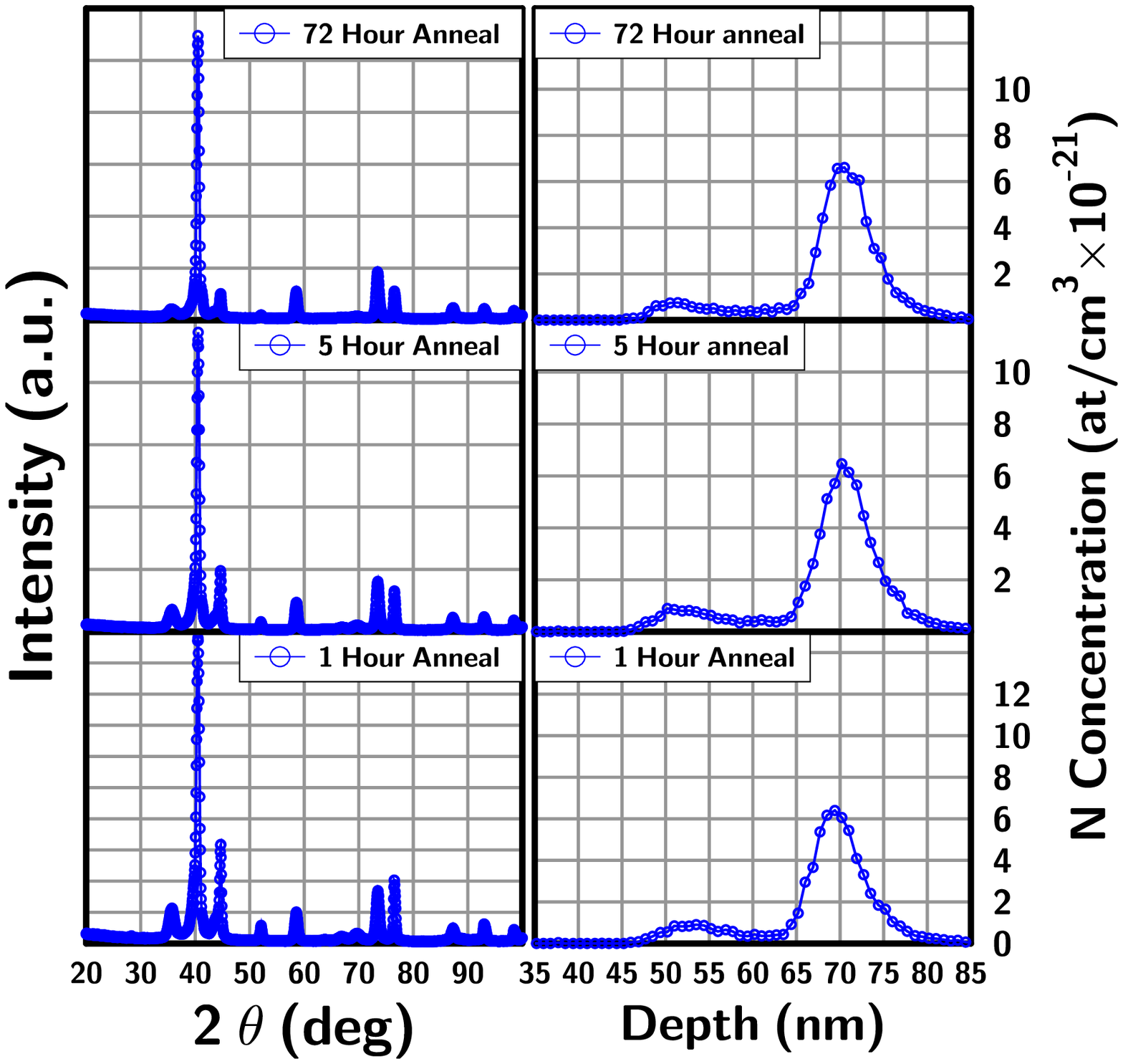}
\caption{(Color online) Grazing incidence \ac{XRD} (Left) and \ac{SIMS} (Right) depth profile of \NNW\ film annealed for 1, 5 and 72 hours is plotted.}
\label{f3}
\end{figure}

\begin{figure}[h!]
\includegraphics[width=7.5cm]{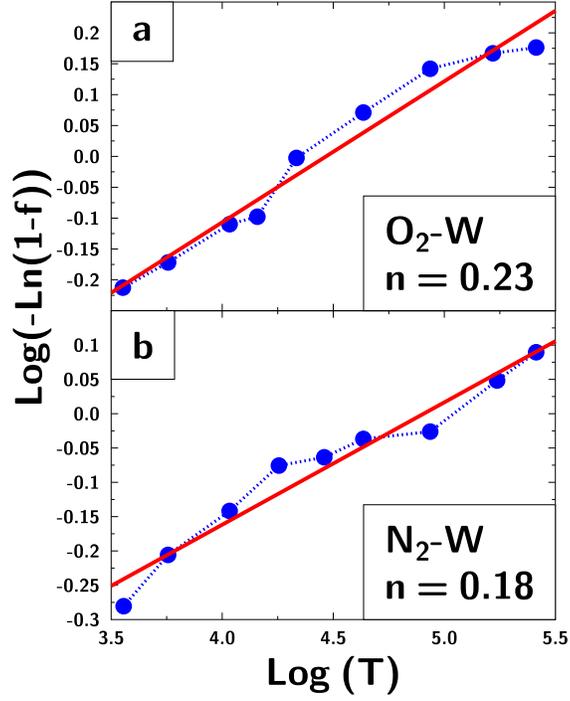}
\caption{(Color online) Linear fit to $\log(-\ln(1-f))$ versus $\log(t)$ is shown, where f is fraction of $\alpha$ phase transformed from $\beta$ and t is annealing time. Growth exponent n is estimated for the \OOW\ and \NNW\ films. }
\label{f4}
\end{figure}

Grazing incidence \ac{XRD} and \ac{SIMS} depth profile of \OOW\ film annealed at a temperature of 175$^\circ$C for 1, 5 and 72 hours is shown in Fig.~\ref{f2}. A clear reduction in $\beta$ W (200) peaks is observed in the \ac{XRD} patterns as the annealing time increases. The tungsten layer in the annealed films is a mixture of $\alpha$ and $\beta$ phase. The relative $\alpha$ phase fraction is 0.45, 0.55 and 0.77 for the 1 hour, 5 hours and 72 hours annealed \OOW\ films. The oxygen concentration profile displays a gradient from the Ni-W (50~nm) and W-Si (70~nm) interface. The O signal in this region decreases at a rate of 28.7 nm/dec (corresponding to 1/e  decay length of 227~\AA) average for all the annealing times. The rate of decrease ($= \frac{\partial C}{\partial t} \propto \frac{\partial^2C}{\partial x^2}  \approx 0$, where $C$ is concentration, $x$ is position and $t$ is time) remains relatively constant for all films annealed at different time lengths. Grazing incidence \ac{XRD} and \ac{SIMS} depth profile of \NNW\ film annealed for 1, 5 and 72 hours are displayed in Fig.~\ref{f3}. The $\alpha$ phase fraction is 0.40, 0.57 and 0.85 for the 1 hour, 5 hours and 72 hours annealed \NNW\ films. The nitrogen concentration profile is similar, irrespective of annealing time lengths. The nitrogen profile in the tungsten layer ($\approx$ 50 nm to 70 nm) decreased at a rate of 31.5 nm/dec (corresponding to decay length 1/e of 279~\AA), averaged for all the annealing times. A plot of $\log(-\ln(1-f))$ versus $\log(T)$ , where f is fraction of $\alpha$ phase formed is shown in Fig.~\ref{f4}. A linear fit to Avrami growth model is also shown as a straight line. A growth exponent of 0.23 for the \OOW\ and 0.18 for the \NNW\ film is estimated in Fig.~\ref{f4}a and b respectively. The fit has a R$^2$ value of about 0.96.


In the \ac{XRD} patterns of the \pureW\ film the Ni \ac{FCC} peaks do not overlap with any of the $\alpha$ phase tungsten peaks. The $\beta$ W(112) peak overlaps with  Ni \ac{FCC} peak around 45$^\circ$ as seen in both the \OOW\ and \NNW\ X-ray scans as indicated in Fig.~\ref{f1}a.  A Rietveld refinement was performed on these patterns to conclude that 2 sccm of O$_2$ gas and 2 second N$_2$ pulse during the deposition of the tungsten both produce a nearly complete $\beta$ phase tungsten film and is consistent with our previous findings~\cite{narasimham:aipa:4:117139,narasimham:aipa:5:117107}.  Oxygen \ac{SIMS} depth profile of 20~nm \pureW\ was barely detectable at 0.03 at.\%, which is close to the instrument background, and therefore there is negligible oxygen in the film. This is in contrast to the 20~nm \OOW\ $\beta$ phase film, which has an average value of about $9.0\pm0.9$~at.\% of oxygen in the tungsten layer. In the as-deposited state of \OOW\ film oxygen atoms in the tungsten layer segregate towards interfaces due to concentration gradient induced diffusion.

The nickel layer is deposited on the tungsten layer, and the latter layer has doped-oxygen. Oxygen atoms diffuse towards nickel due to concentration gradient ($\frac{\partial C}{\partial x} \neq 0$). In the \pureW\ film no oxygen is present in the tungsten layer. The peak in oxygen profile near the W-Si interface is due to the native oxide on Si. Comparing the \pureW\ and \OOW\ oxygen profiles around the W-Si interface, it is evident that the doped-oxygen in the \OOW\ film diffuses towards the W-Si interface. The chemical bonding between Si-O can also act as driving force for this diffusion. The dip at this interface is due to W-Si bonds which reduce sites for oxygen bonding and also presence of silicon stabilizes the $\beta$ phase~\cite{jansson:ass:73:51}. The oxygen profile reaches a steady-state ($\frac{\partial C}{\partial x} \cong 0$) in the tungsten layer, as the oxygen atoms gets trapped in the interstitial sites to form stable $\beta$ phase tungsten as confirmed from XRD~\cite{sluiter:prb:80:220102}.
The \NNW\ film also crystallizes in 100~\% $\beta$ phase but, shows a drastic reduction of dopant concentration to $0.15\pm0.02$ at.\% of nitrogen in the tungsten layer when compared to oxygen in the \OOW\ film. This value is much lower than the theoretically predicted value of 10~at.\%~\cite{sluiter:prb:80:220102}. The lower electronegativity (relative to oxygen) of nitrogen and absence of any excess nitrogen atoms during the growth can be attributed to such a low concentration. The sudden increase in the nitrogen concentration profile at the W-Si interface is due to the 5 nm SiN layer.

The annealing temperature used is less than one-third the melting point of nickel and tungsten. Thus, no significant self-diffusion or interdiffusion of nickel and tungsten will occur at an annealing temperature of 175$^\circ$C. The annealed \OOW\  and \NNW\ films have a phase mixture of $\alpha$ and $\beta$, with the phase fraction skewed towards $\alpha$. Alpha W has 2 atoms per unit cell and $\beta$ W has 8 atoms per unit cell~\cite{hagg:acb:7:351}. The $\beta\rightarrow\alpha$ transformation increase partially unsaturated tungsten atoms at the interfaces. The tungsten layer shrinks in size by $\approx$ 4.5~nm, due to smaller lattice parameter of $\alpha$ phase than $\beta$ phase. As the \OOW\ film is annealed at 175$^\circ$C, oxygen atoms diffuse within the film. The diffusion of oxygen atoms is mainly driven by thermal energy and the preexisting concentration gradient. As the tungsten layer phase transforms to $\alpha$, the oxygen (nitrogen) atoms can either diffuse to interfaces or dissolve in the interstitial sites of $\alpha$ W. Thus, limiting the diffusion of oxygen atoms between the W-Ni and W-Si interface and no steady-state concentration of oxygen is observed. As the W layer is deposited onto the Si substrate, the oxygen atoms would diffuse towards the partially saturated Si substrate during the growth. Once the $\beta$ W is fully deposited, the oxygen atoms are mainly trapped in the interstitial sites in the tungsten layer. The  excess atoms diffuse towards the Ni interface, thus when the phase transformation occurs, the newly available oxygen atoms diffuse towards the Ni interface.

In case of \NNW\ films, the nitrogen atoms present, in the tungsten layer, are not in excess. As the film is annealed, nitrogen atoms diffuse due to concentration gradient. Excess nitrogen atoms at the interface of W-SiN diffuses into the tungsten layer. No significant change in the nitrogen profile is observed for the samples annealed at different time lengths. The growth exponent of phase transformation of $\beta\rightarrow\alpha$ for both \OOW\ and \NNW\ is 0.23 and 0.18 respectively, seen in Fig.~\ref{f4}.  This indicates that the transformation is initiated by non-uniform nucleation of $\alpha$ phase, which grow at the expense of $\beta$ phase. However, there are possible mechanisms which depend on the strength and nature of correlation of the nuclei, which generate a growth exponent less than 0.5, but more analysis is required to arrive at a deeper conclusion~\cite{rickman:pre:95:022121,tong:jcp:114:915,rickman:acma:45:1153}.

The rate of decreases of dopant concentration and the phase fraction do not seem to correlate proportionally with one other. The  \OOW\ and \NNW\  film samples annealed for 72~hours are predominantly $\alpha$ phase with about 11 at.\% and 14 at. \% of $\beta$ phase respectively. The oxygen or nitrogen atoms still present in the tungsten layer can stabilize a fraction of tungsten layer in $\beta$ phase by dissolution of dopants in interstitial sites~\cite{sluiter:prb:80:220102}. In the as-deposited state of \OOW\ film, the $\beta$ phase can be stabilized by abundantly available oxygen atoms. In case of the \NNW\ film it could be a template assisted-growth as seen in other studies~\cite{liu:acma:104:223}. This observation of a highly crystalline $\beta$ W with such a low dopant concentration is a surprising result and is much lower than what has been predicted by \emph{ab initio} calculations~\cite{sluiter:prb:80:220102}.


The $\beta$ phase is a closed packed structure with a lower specific surface free energy than the $\alpha$ phase, which would promote the formation of $\beta$ tungsten in the initial stages of deposition~\cite{markov:book:1}.  The addition of dopants helps to further stabilize this phase. The theoretically calculated stabilization energy for $\beta$ (A15) W is 1 eV per O atom, which is much higher than the available thermal energy at room temperature (25.7 meV) and even at 175$^\circ$C (38.6 meV). However, the $\beta$ to $\alpha$ phase transformation is observed with such a low thermal energy, suggesting a different stabilization mechanism of $\beta$ tungsten, such as grain size restriction from dopant atoms or template assisted growth, which are not accounted for in \emph{ab initio} calculations.



In conclusion, the \NNW\ forms a $\beta$  phase film with a much lower dopant concentration of 0.15~at.\% of nitrogen than when compared to the \OOW\ film (9 at.\% of oxygen). The pulsing of N$_2$ gas can be utilized to obtain ultra pure films of $\beta$ phase of W, which would be useful in uncovering the underlying physical mechanisms of the \ac{GSHE}. The phase transformation of \OOW\ and \NNW\ from 100~\% $\beta$ phase to predominantly $\alpha$ is triggered by non-uniform $\alpha$ nuclei. No proportional correlation between phase fraction and dopant concentration was observed, suggesting either an independent degree of stabilization or limiting degree of stabilization by dopants for $\beta$ W.

\section{Acknowledgements}

The authors gratefully acknowledge the support by the Nanoelectronics Research Initiative (NRI) through the INDEX center under spin logic theme ID:2399.002 and the National Science Foundation (DMR-1308102). A special thanks to Kalaga Kaushik for facilitating Rietveld refinements studies.

\pagebreak

\bibliography{iiiv}

\pagebreak

\

\pagebreak
\

\pagebreak
\

\pagebreak

\end{document}